\documentclass[preprint,showpacs,preprintnumbers,amsmath,amssymb]{revtex4}

\usepackage{dcolumn}
\usepackage{bm}
\usepackage[dvipdfmx]{graphicx}
\usepackage{color}

\begin{document}

\preprint{}

\title{Accelerated diffusion by chaotic fluctuation in probability in photoexcitation transfer system}

\author{Song-Ju Kim$^{1}$}
\email{KIM.Songju@nims.go.jp}

\author{Makoto Naruse$^{2}$}

\author{Masashi Aono$^{3}$}

\author{Hirokazu Hori$^{4}$}

\author{Takuma Akimoto$^{5}$}
\email{akimoto@keio.jp}

\affiliation{%
$^{1}$WPI Center for MANA, National Institute for Materials Science, Tsukuba, Ibaraki 305--0044, Japan
}%

\affiliation{%
$^{2}$PNRI, National Institute of Information and Communications Technology, Tokyo 184--8795, Japan
}%

\affiliation{%
$^{3}$Earth-Life Science Institute, Tokyo Institute of Technology, Tokyo 152--8550 \& PRESTO JST, Japan
}%

\affiliation{%
$^{4}$Graduate School of Medicine and Engineering, University of Yamanashi, Yamanashi 400-8511, Japan
}%

\affiliation{%
  $^{5}$Department of Mechanical Engineering, Keio University, Kohoku-ku, Yokohama 223-8522, Japan
}%

\date{\today}

\begin{abstract}
We report a new accelerated diffusion phenomenon that is produced by a one-dimensional random walk in which the flight probability to one of the two directions (i.e., bias) oscillates dynamically in periodic, quasiperiodic, and chaotic manners.
The probability oscillation dynamics can be physically observed in nanoscale photoexcitation transfer in a quantum-dot network, where the existence probability of an exciton at the bottom energy level of a quantum dot fluctuates differently with a parameter setting.
We evaluate the ensemble average of the time-averaged mean square displacement (ETMSD) of the time series obtained from the quantum-dot network model that generates various oscillatory behaviors because the ETMSD exhibits characteristic changes depending on the fluctuating bias; in the case of normal diffusion, the asymptotic behavior of the ETMSD is proportional to the time (i.e., a linear growth function), whereas it grows nonlinearly with an exponent greater than $1$ in the case of superdiffusion.
We find that the diffusion can be accelerated significantly when the fluctuating bias is characterized as weak chaos owing to the transient nonstationarity of its biases, in which the spectrum contains high power at low frequencies.
By introducing a simplified model of our random walk, which exhibits superdiffusion as well as normal diffusion, we explain the mechanism of the accelerated diffusion by analyzing the mean square displacement.
\end{abstract}

\maketitle

\section{Introduction}
Previously, the authors theoretically and numerically demonstrated that chaotic oscillation occurs in a nanoscale system consisting of a pair of quantum dots (QDs), between which energy transfers via optical near-field interactions~\cite{nanochaos}.
The chaotic oscillation occurs at the values of the existence probability of an exciton in one of the QDs when the QDs are connected with an external time delay.
It is scientifically and technologically important to elucidate the exact nature of this new oscillatory phenomenon, which we call ``nanochaos,'' since it implies that ultrafast random number generators~\cite{nanochaos}, problem solvers~\cite{PRB,Lang}, and decision makers~\cite{QDM,QDM2,QDM3} can be constructed by the nanoscale elements.
For simplicity in modeling the oscillation in the probability of random selection operations, we consider a one-dimensional random-walk model in this study, in which the flight probability to one of the two directions, which is called ``bias,'' fluctuates.

Our random-walk model with a fluctuating bias is different from the previously studied Langevin equation model that has a fluctuating diffusion coefficient, which was proposed to investigate the diffusion phenomena of a particle with inner degrees of freedom, e.g., a cell, the center-of-mass motion in the reptation model, and diffusion in a heterogeneous environment~\cite{Massignan2015,Manzo2015,Uneyama2015,Yamamoto2015,Akimoto-Seki2015}.
In fact, the diffusivity in our model does not change, even while the bias fluctuates.
On the other hand, Sinai's random-walk model uses a bias that fluctuates depending on the position of a particle~\cite{sinai,sinai2}.
Our model is used to evaluate how the chaotic dynamics in the time-dependent bias fluctuation contribute to diffusion phenomena, whereas the Sinai model having the position-dependent bias is not useful for our purposes.

In Section II, we will define our random-walk model with a fluctuating bias by introducing some time series data taken from numerical simulations of the quantum-dot network model.
In Section III, we will present simulation results obtained by our model, including a new accelerated diffusion phenomenon.
To clarify the mechanism of accelerated diffusion, we will introduce a simplified model of our random walk and will present theoretical explanations of the mechanism in Section IV.
We will conclude the paper by providing some discussion of the implications of our results.

\section{Methods}

Random walks can be generated by chaotic time series. In previous studies, we usually obtain coarse-grained (binary) time series of the chaotic times series by using a fixed threshold of $0.5$~\cite{chaos}.
In other words, we use chaotic time series instead of a random number generator (RNG), where a right flight is generated ($X(t)$ $=$ $+1$) if a random number $r(t)$ $>$ $0.5$ or a left flight ($X(t)$ $=$ $-1$) otherwise.
The trajectory $x(t)$ $=$ $X(1)$ $+$ $\cdots$ $+$ $X(t)$ represents the position of the random walker at time $t$.

\begin{figure}[t]
\centering
\includegraphics[height=80mm]{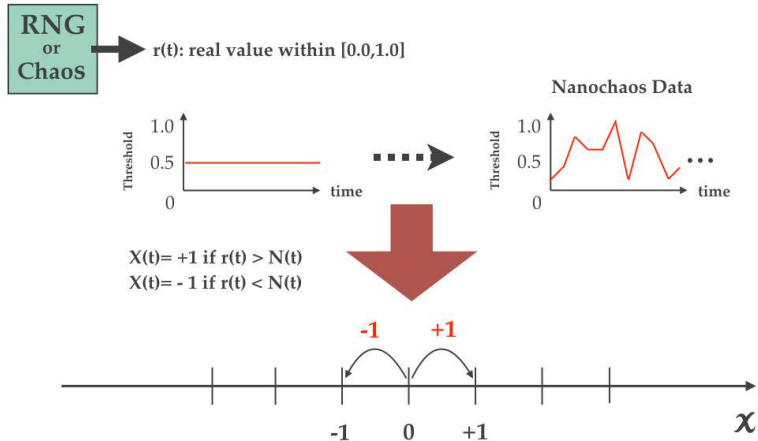}
\caption{Method for treating chaotic time series as probabilities. }
\label{fig:1}
\end{figure}

On the other hand, one can use chaotic time series as threshold time series.
In other words, we treat chaotic time series as the ``probabilities'' for a left flight, as shown in Fig.~\ref{fig:1}.
Therefore, the bias fluctuates in this random walk.

\begin{figure}[t]
\centering
\includegraphics[height=100mm]{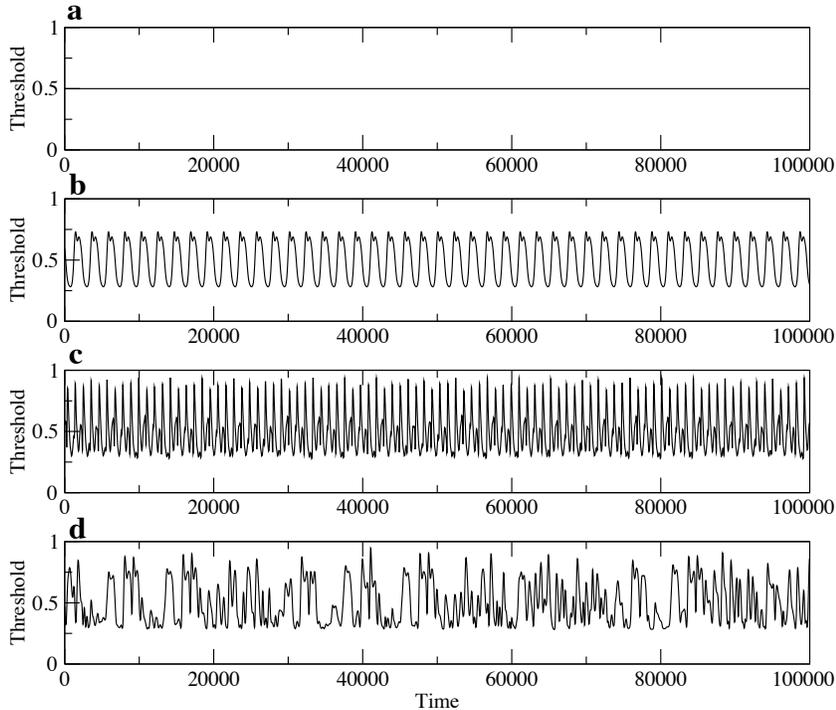}
\caption{Normalized threshold time series: (a) fixed, (b) periodic (No. 30), (c) quasiperiodic (No. 207), and (d) chaotic (No. 180) time series. Here, the numbers denote the parameter numbers used in Ref.~\cite{nanochaos}. }
\label{fig:2}
\end{figure}
First, we prepare a good RNG that generates real values between $0.0$ and $1.0$.
We used the Mersenne Twister (MT)~\cite{MT} in this study.
Using the RNG, we treat each value of the time series as a threshold at the time.
In other words, if the value of the RNG at time $t$ is larger than the threshold (the value of the time series), $X(t)$ $=$ $+1$ (right flight); otherwise, $X(t)$ $=$ $-1$ (left flight).
To construct a normalized threshold time series within $[0.0,1.0]$, we executed the following preprocessing for the time series we use, $S(t)$ ($t = 1$, $\cdots$, $T$):
\begin{equation}
 N(t) = \frac{S(t) - m}{6 \sigma} + 0.5,
\label{pre}
\end{equation}
where $m$ and $\sigma$ are the mean and standard deviation of $S(t)$, respectively.
Thus, the mean of $N(t)$ is exactly $0.5$.

As a result, we can obtain the normalized threshold (probability) time series $\{$ $N_1(t)$, $N_2(t)$, $N_3(t)$, $N_4(t)$  $\}$, as shown in Fig.~\ref{fig:2}.
We prepare four representative time series from a data set of nanophotonic oscillations~\cite{nanochaos}.
Figure~\ref{fig:2} shows (a) fixed, (b) periodic (No. 30), (c) quasiperiodic (No. 207), and (d) chaotic (No. 180) time series.
In what follows, we used these time series with $T$$=$$1,000,000$, although only $100,000$ time steps~\footnote{Actually, owing to the transition period, we used normalized time steps ($T$) of (a) $1,000,000$, (b) $990,000$, (c) $900,000$, and (d) $900,000$.} are shown in Fig.~\ref{fig:2}.
Here, the numbers denote the parameter number used in Ref.~\cite{nanochaos}.

\section{Simulation Results}

Diffusion can be characterized by the mean square displacement (MSD). There are two procedures for obtaining the average of the MSD, i.e., the ensemble average and time average.
Although one can use different realizations of the fluctuating bias in the random walk to calculate the MSD, we take the average for a single time series of the bias generated by the previous study~\cite{nanochaos}.
It follows that the MSD crucially depends on the time series. To extract overall diffusion property, we take both averages, i.e.,
the ensemble average of the time-averaged mean square displacement (ETMSD), defined as
\begin{equation}
{\rm ETMSD}(\tau) = \langle \frac{1}{T-\tau} \sum_{t=1}^{T-\tau} ( x(t + \tau) - x(t) )^2    \rangle ,
\label{eq1}
\end{equation}
where $x(t)$ is the position at time $t$ with the initial condition $x(0)=0$, $\langle \cdots \rangle$ denotes the ensemble average with respect to the noise (the seed of the RNG), and $T$ is the total measurement time length ($1,000,000$). We note that the ensemble average is taken under one realization of the fluctuating bias (the threshold time series is fixed). When there is no bias, i.e., $N(t)=0.5$, the ETMSD exhibits normal diffusion, i.e., ETMSD($\tau$) $= \tau$.

\begin{figure}[t]
\centering
\includegraphics[width=120mm]{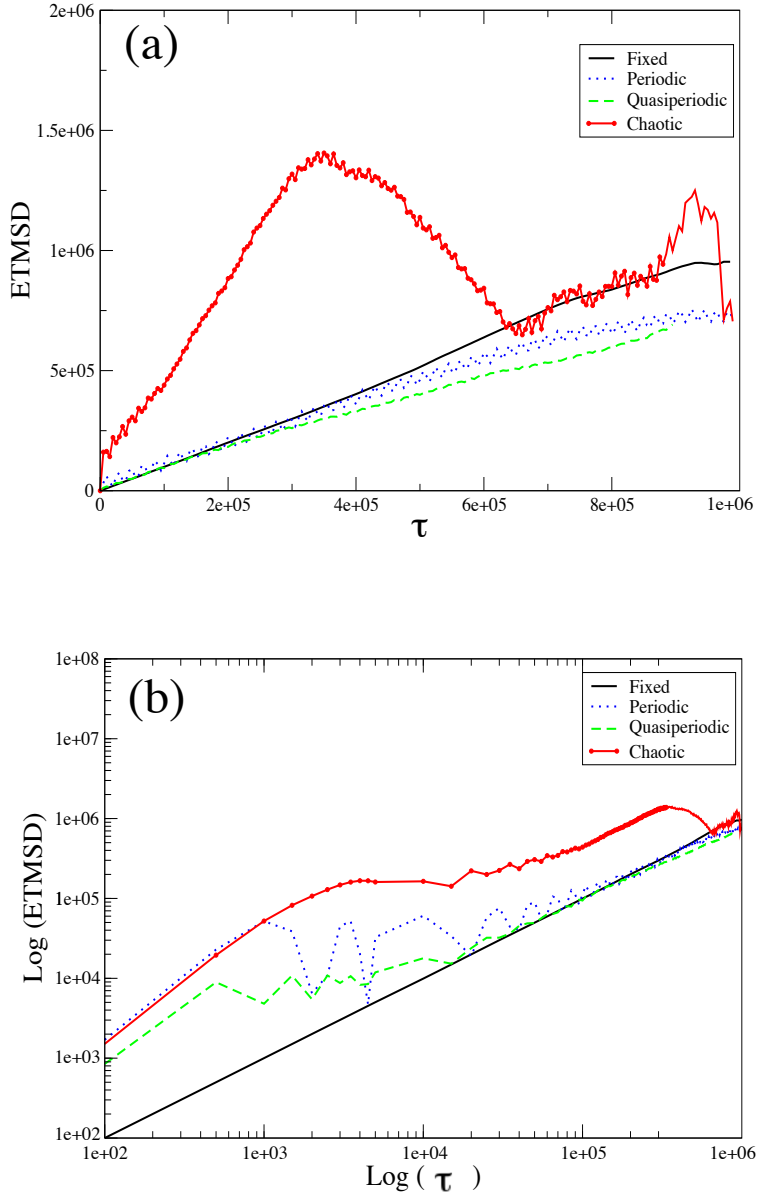}
\caption{(a) ETMSDs for fixed (black solid line), periodic (blue dotted line), quasiperiodic (green dashed line), and chaotic (red filled circles) probabilities. (b) Fig.~\ref{fig:3}(a) plotted with logarithmic scales.}
\label{fig:3}
\end{figure}
Figures~\ref{fig:3}(a) and (b) show the ETMSDs of the time series shown in Fig.~\ref{fig:2}.
We can confirm that the ETMSDs do not monotonically increase, except in the ordinary one (the case with a fixed probability of $0.5$).
In particular, the ETMSD for the chaotic time series undergoes many large changes in a short time interval owing to the transient nonstationarity of its biases, although the overall time dependence of the ETMSD is proportional to the time, which indicates normal diffusion, as shown in Fig.~\ref{fig:3}(b).
We confirmed that the ETMSD at $\tau=10^6$, $2 \times 10^6$, $\cdots$ increases linearly with the time
when we periodically generate $N(t)$ with a period of $10^6$ (not shown).

\begin{figure}[t]
\vspace{3mm}
\centering
\includegraphics[height=80mm]{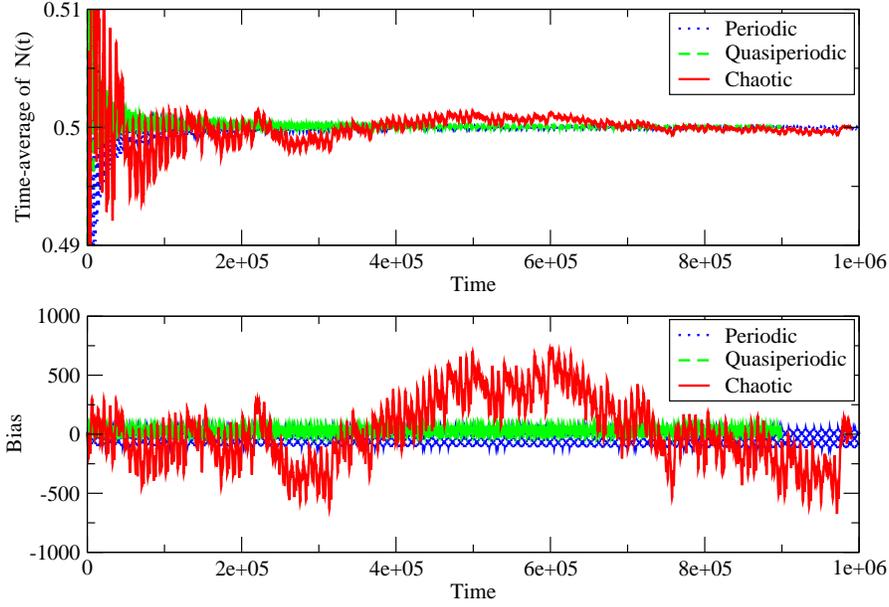}
\caption{(a) The time average of $N(t)$ for periodic (blue dotted line), quasiperiodic (green dashed line), and chaotic (red solid line) probabilities. (b) The average difference between the number of occurrences of $1$ and the number of occurrences of $0$ until time $t$.}
\label{conv}
\end{figure}

This nonstationarity originates from the bias of $0$ or $1$ in a local time interval, as shown in Fig.~\ref{conv}(b).
This is because the spectrum of the chaotic oscillation (No. 180) contains high power at low frequencies.
It is noted that the ETMSD of a strongly chaotic data such as a logistic map or the No. 180 chaotic data with a destructive time correlation (surrogate data) exhibits similar behavior to that of the fixed threshold (monotonically increasing).
There exists a huge bias in a short time interval, although these biases are canceled in the overall time series owing to the normalization (see Section~II), as shown in Fig.~\ref{conv}(a).

\begin{figure}[t]
\centering
\includegraphics[height=80mm]{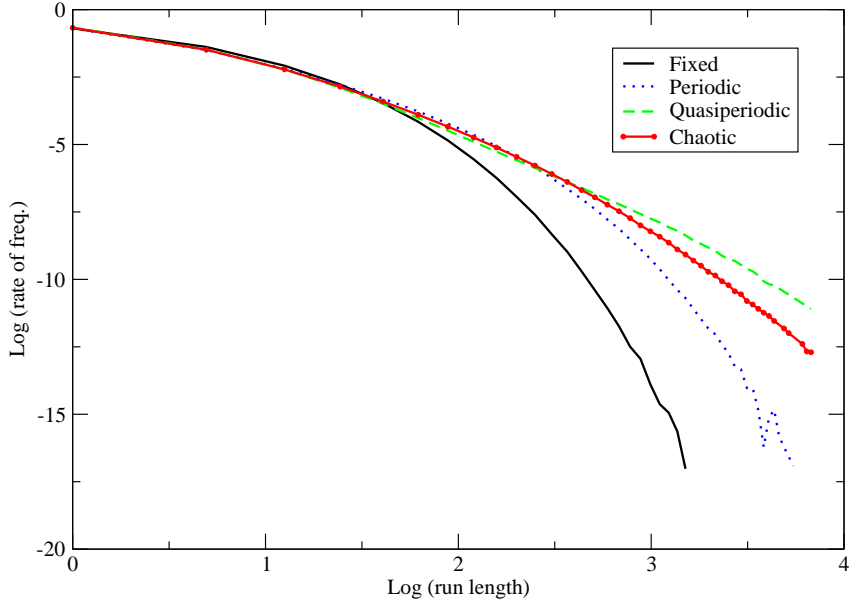}
\caption{The sojourn-time distributions of $X(t)$ generated from the time series of fixed (black solid line), periodic (blue dotted line), quasiperiodic (green dashed line), and chaotic (red filled circles) probabilities.}
\label{fig:5}
\end{figure}
Figure~\ref{fig:5} shows the sojourn-time distributions of $X(t)$ generated from the threshold time series shown in Fig.~\ref{fig:2}.
The log-log scale distributions are shown for fixed (black solid line), periodic (blue dotted line), quasiperiodic (green dashed line), and chaotic (red filled circles) probabilities.
We can confirm that the sojourn-time distributions come close to the power law distribution as the time series become chaotic.
The power law distribution means that there is no characteristic sojourn-time length.
This is one of the properties of superdiffusion shown in the L\'evy walk~\cite{Schlesinger1982,Geisei}.
However, the power law distribution is not perfect in the case of No. 180.
The $1$-run distribution does not follow a power law, whereas the $0$-run distribution follows a short power law.
As a result, the overall property is classified as normal diffusion.

Among the nanochaos data sets, we can observe superdiffusion in our model if we find a time series of chaotic probability, such as the modified Bernoulli map (Aizawa map~\cite{chaos,aizawa}), that generates power law distributions.
The Aizawa map is described by following equations:
\begin{equation}
S(t+1) = \left\{
\begin{array}{ll}
S(t) + 2^{B-1} S(t)^B \hspace{15mm}(0  \leq S(t) \leq 0.5)\\
\\
S(t) - 2^{B-1} (1 - S(t))^B \hspace{5mm}(0.5 < S(t) \leq 1).
\end{array}
\right.
\end{equation}

\begin{figure}[t]
\centering
\includegraphics[height=90mm]{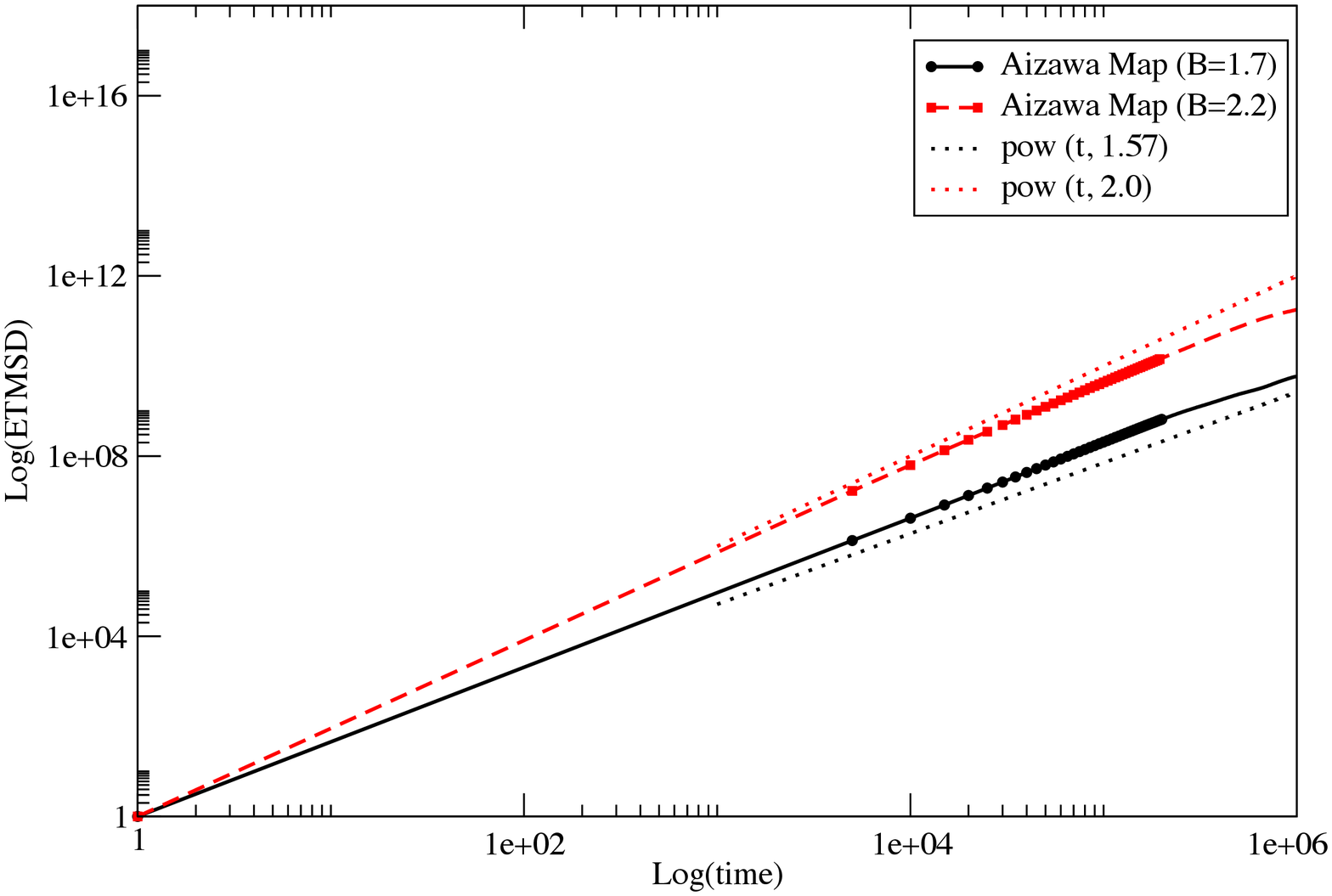}
\caption{ETMSDs of a random walk using a time series generated by an Aizawa map with $B=1.7$ (black circles) and $B=2.2$ (red squares).
The slope of each ETMSD indicates superdiffusion, which is very close to the expected value ($1.57$ or $2.0$) from our theory (see Section~\ref{ST}).}
\label{fig:6}
\end{figure}

Figure~\ref{fig:6} shows the ETMSDs of a random walk using a time series generated by an Aizawa map with $B=1.7$ and $B=2.2$.
Here, for the purposes of correctly estimating the slope, we do not use normalization (Eq.~(\ref{pre})) and use $100$ samples from different initial conditions.

The time dependences of the MSD ($=$$t^{\beta}$) for the Aizawa map, which has one parameter $B$, are as follows:
\begin{itemize}
\item
$B \leq 1.5$: normal diffusion $(\beta = 1.0)$,
\item
$1.5 < B \leq 2.0$: superdiffusion $(\beta = 3 - \frac{1}{B-1})$,
\item
$B > 2.0$: superdiffusion $(\beta =2.0)$.
\end{itemize}
In next section, we show analytical calculations of these MSDs using a simplified stochastic model. 
Each of these results corresponds to Eq. (\ref{t1.0}), Eq. (\ref{t1.7}), and Eq. (\ref{t2.0}), respectively, in the next section.
In Fig.~\ref{fig:6}, we can confirm that the exponent $\beta$ indicates superdiffusion, which is very close to the expected value $1.57$ or $2.0$ from our theory (see section~\ref{ST}).


\section{Theoretical Results}
\label{ST}

\begin{figure}[t]
\centering
\includegraphics[height=80mm]{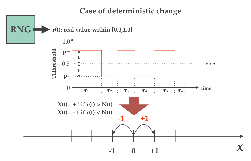}
\caption{Simplified stochastic model (deterministic change).}
\label{fig:T}
\end{figure}

In this section, we consider a simple stochastic model related to a random walk with a fluctuating bias, which can generate a variety of diffusion types, as shown in Fig.~\ref{fig:T}.
The model is a simple generalization of a L\'evy walk, in which a random walker performs a two-state biased random walk with a random persistence time.
The probability of going to the left is given by $p_\pm=0.5 \pm \epsilon$ ($\epsilon=0.5$ corresponds to the ordinary L\'evy walk),
where the subscript represents the state, i.e., the $+$ or $-$ state.
The MSDs can be analytically calculated for several persistence time distributions.
Using the analytical results of the stochastic model, we can understand that the overall properties in nanochaos data we treat in this paper are all classified as normal diffusion (see Eq.~(\ref{t1.0})).

\subsection{Stochastic model}
In a L\'evy walk, a random walker moves to the right or left with a constant velocity for a continuous random persistence time. After the persistence time, the random walker can change direction.
Instead, a random walker in our model performs a biased random walk with a probability of going to the left of $p_\pm=0.5 \pm \epsilon$ (the probability of going to the right is $q_\pm=1-p_\pm$) for a continuous random persistence time. In what follows, we consider the continuous version of the random walk.
Thus, we use the following propagator during the biased random walk with the state $\pm$:
\begin{equation}
P_\pm(x,t) = \frac{1}{\sqrt{4\pi Dt}} e^{-\frac{(x\mp c t)^2}{4Dt}}.
\end{equation}
We note that the diffusion coefficient $D$ and the velocity $c$ are expressed as $2D=4p_\pm q_\pm = 1 - {4} \epsilon^2$ and
$c=p_+- q_+= 2 \epsilon$, respectively.
We use $\rho(t)$ as the probability density function (PDF) of the persistence times.

\subsection{Random walk framework}


Let $Q_{\pm}(x,t)$ be the joint PDF of finding a random walker at position $x$ at time $t$, and the states changes from $\mp$ to $\pm$ exactly at $t$. Further, let $R_{\pm}(x,t)$ be the PDF of finding a random walker with the state $\pm$ at position $x$ at time $t$. Then, we have the following Montroll--Weiss equations (deterministic change of state)~\cite{Schlesinger1982,mont}:
\begin{equation}
Q_{\pm}(x,t) = p_{\pm}^{0} \delta(x)\delta(t) + \int_0^t dt' \int_{-\infty}^{\infty} dx' \psi_\mp (x',t') Q_\mp (x-x',t-t')
\label{qpm}
\end{equation}
and
\begin{equation}
R_{\pm}(x,t) = \int_0^t dt' \int_{-\infty}^{\infty} dx' \Psi_\pm (x',t') Q_\pm (x-x',t-t'),
\label{rpm}
\end{equation}
where $p_{\pm}^{0}$ is the probability of the initial state ($\pm$),
$\psi_\pm (x,t)= P_\pm (x,t) \rho(t)$, and $\Psi_\pm (x,t)= P_\pm (x,t) \int_t^\infty dt' \rho(t')$.
Note that the initial position is $x=0$, and we assume that the initial persistence time distribution
is the same as $\rho(t)$.
Using the Fourier--Laplace transform of $Q_\pm(x,t)$ with respect to $x$ and $t$ in Eq.~(\ref{qpm}), we have $\hat{Q}_\pm(k,s)$.
Then, the substitution of this into the Fourier--Laplace transform of $\hat{R}_\pm(k,s)$ in Eq.~(\ref{rpm}) gives
\begin{equation}
\hat{R}_{\pm}(k,s) = \frac{p_{\pm}^{0} + p_{\mp}^{0} \hat{\psi}_\mp (k,s)}{1 - \hat{\psi}_\pm (k,s) \hat{\psi}_\mp (k,s)}
\hat{\Psi}_{\pm} (k,s).
\end{equation}
When $p_{\pm}^{0} = 1/2$, we have
\begin{equation}
\hat{R}_{\pm}(k,s) = \frac{1}{2}\frac{1 +  \hat{\psi}_\mp (k,s)}{1 - \hat{\psi}_\pm (k,s) \hat{\psi}_\mp (k,s)} \hat{\Psi}_{\pm} (k,s),
\end{equation}
and {$R(x,t) \equiv R_+(x,t) + R_-(x,t)$} becomes
\begin{equation}
\hat{R}(k,s) = \frac{\hat{\Psi} (k,s) +  \hat{\psi \Psi} (k,s)}{1 - \hat{\psi}_\pm (k,s) \hat{\psi}_\mp (k,s)},
\end{equation}
where $\hat{\Psi}(k,s)= \{\hat{\Psi}_+(k,s) + \hat{\Psi}_-(k,s) \}/2$, and
$\hat{\psi\Psi}(k,s)= \{\hat{\psi}_+(k,s)\hat{\Psi}_-(k,s) + \hat{\psi}_-(k,s)\hat{\Psi}_+(k,s) \}/2$.
We note that $\hat{\psi}_\pm (0,s) = \hat{\rho}(s)$, $\hat{\Psi}_\pm (0,s) = \{1-\hat{\rho}(s)\}/s$,
$\hat{\Psi} (0,s) = \{1-\hat{\rho}(s)\}/s$, and $\hat{\psi\Psi} (0,s) = \hat{\rho}(s) \{1-\hat{\rho}(s)\}/s$.

\subsection{Persistence time distribution}

\subsubsection{Deterministic case (periodic)}
Here, we consider that the persistence time of the state is constant ($\tau_0$): $\rho(t) = \delta (t - \tau_0)$.
By Eq.~(\ref{qpm}), we have
\begin{equation}
Q_{\pm}(x,\tau_0) = \frac{P_\mp (x,\tau_0)}{2}
\end{equation}
and
\begin{equation}
Q_{\pm}(x,2\tau_0) = \int_{-\infty}^\infty dx' \psi_\mp (x',\tau_0) Q_\mp (x-x',\tau_0).
\end{equation}
The Fourier transform with respect to $x$ is
\begin{equation}
\tilde{Q}_{\pm}(k,2n\tau_0+\tau_0) = \frac{\{\hat{\psi}_+ (k,\tau_0) \hat{\psi}_-(k,\tau_0)\}^n \hat{\psi}_\pm(k,\tau_0)}{2}
\end{equation}
and
\begin{equation}
\tilde{Q}_{\pm}(k,2n\tau_0) = \frac{\{\hat{\psi}_+ (k,\tau_0) \hat{\psi}_-(k,\tau_0)\}^n}{2}
\end{equation}
for $n=0,1, \cdots$.

\subsubsection{Stochastic case}
Here, we consider three cases for the PDFs of the persistence times whose Laplace
transforms are given by
  \begin{itemize}
    \setlength{\leftskip}{4.mm}
    \item [(1)] $\alpha =2$: \hspace*{.9cm}
    $\hat{\rho}(s) = 1 - \mu s + as^{2} + o(s^{\alpha})$,\\[-.2cm]
    \item [(2)] $1 <\alpha <2$: \quad
    $\hat{\rho}(s) = 1 - \mu s + a s^{\alpha} + o(s^{\alpha})$,\\[-.2cm]
    \item [(3)] $\alpha<1 $: \hspace*{.85cm}
    $\hat{\rho}(s) = 1 - a s^{\alpha} + o(s^{\alpha})$,
  \end{itemize}
where $\mu$ is the mean persistence time and $a$ is an arbitrary value. The mean persistence time diverges for case (3), and
the second moment of the persistence time diverges for cases (2) and (3).
For example, cases (2) and (3) correspond to
\begin{equation}
\int_t^\infty dt' \rho(t') \sim \left(\frac{c_0}{t}\right)^\alpha\quad(t\to\infty),
\end{equation}
where $c_0$ is a microscopic scale parameter.
Because
\begin{equation}
\hat{\psi} (k,s) = \int_{-\infty}^\infty dx \int_0^\infty dt e^{-st -kx} \psi (x,t),
\end{equation}
the derivative with respect to $k$ gives
\begin{equation}
\hat{\psi}' (k,s) = \int_{-\infty}^\infty dx \int_0^\infty dt (-x) e^{-st -kx} \psi (x,t),
\end{equation}
implying $\hat{\psi}' (0,s)=0$. Moreover,
\begin{equation}
\hat{\psi}'' (0,s) = \int_0^\infty dt e^{-st} (2Dt + c^2 t^2) \rho(t).
\end{equation}
For $\alpha < 2$,
\begin{equation}
\hat{\psi}'' (0,s) \sim c^2 \int_0^\infty dt e^{-st} t^2 \rho(t) \quad (s\to 0).
\end{equation}
In a similar manner, we have $\hat{\Psi}' (0,s)=0$ and
\begin{equation}
\hat{\Psi}'' (0,s) = \int_0^\infty e^{-st} (2Dt + c^2 t^2) \int_t^\infty dt'\rho(t').
\end{equation}
For $\alpha < 2$,
\begin{equation}
\hat{\Psi}'' (0,s) \sim c^2 \int_0^\infty e^{-st} t^2 \int_t^\infty dt'\rho(t') \quad (s\to 0).
\end{equation}

\subsection{Mean square displacement}

\subsubsection{Deterministic change}

The Laplace transform of the mean displacement is given by
 \begin{equation}
\mathcal{L} (\langle x(t) \rangle)(s) = \left. \frac{\partial \hat{R}(k,s) }{\partial k} \right|_{k=0}
=0.
\end{equation}
The Laplace transform of the MSD is given by
\begin{eqnarray}
\mathcal{L} (\langle x(t)^2 \rangle)(s) &=& \left. \frac{\partial^2 \hat{R}(k,s) }{\partial k^2} \right|_{k=0}\\
&=&\left[ \frac{\hat{\Psi}'' (0,s) + \hat{\psi\Psi}'' (0,s) }{1 - \hat{\rho} (s)^2 }
+ \frac{2}{s}\frac{\hat{\psi}'' (0,s) + \hat{\psi}_+' (0,s) \hat{\psi}_-' (0,s) }{ 1 - \hat{\rho} (s)^2 }  \right].
\end{eqnarray}
For case (1),
\begin{equation}
\mathcal{L} (\langle x(t)^2 \rangle)(s) \sim \frac{2D\mu+ (2a - \mu^2)c^2}{\mu s^2}.
\end{equation}
The inverse transform is
\begin{equation}
\langle x(t)^2 \rangle  \sim  \frac{2D\mu+ (2a - \mu^2)c^2}{\mu }t.
\label{t1.0}
\end{equation}
For case (2),
\begin{equation}
\mathcal{L} (\langle x(t)^2 \rangle)(s) \propto \frac{1}{s^{4-\alpha}}.
\end{equation}
The inverse transform is
\begin{equation}
\langle x(t)^2 \rangle  \propto t^{3-\alpha}.
\label{t1.7}
\end{equation}
For case (3),
\begin{equation}
\mathcal{L} (\langle x(t)^2 \rangle)(s) \propto \frac{1}{s^{3}}.
\end{equation}
The inverse transform is
\begin{equation}
\langle x(t)^2 \rangle  \propto t^{2}.
\label{t2.0}
\end{equation}
Therefore, superdiffusion is observed for cases (2) and (3), which is consistent with the L\'evy walk.

When the persistence time is periodic, the MSD is given by
\begin{eqnarray}
\langle x(n\tau_0)^2 \rangle &=& \left. \frac{\partial^2 \tilde{Q}(k,n\tau_0) }{\partial k^2} \right|_{k=0}.
\end{eqnarray}
Because $\tilde{Q}_{\pm}''(0,2n\tau_0) = 4nD\tau_0$ and $\tilde{Q}_{\pm}''(0,2n\tau_0) = 2D(2n +1)\tau_0 + c^2 \tau_0^2$,
the MSD becomes
\begin{equation}
\langle x(k\tau_0)^2 \rangle = \left\{
\begin{array}{ll}
2D(2n +1)\tau_0 + c^2 \tau_0^2 \quad &(k = 2n +1)\\
\\
4nD\tau_0  &(k=2n).
\end{array}
\right.
\end{equation}
Because $D$ is related to the bias, i.e., $D=1-4\epsilon^2$, the bias decreases
the diffusivity at $t=2n\tau_0$. However, the diffusivity at $t=(2n+1)\tau_0$ is greatly enhanced by
a large period $\tau_0$ and bias $c=2\epsilon$. This is a mechanism of accelerated diffusion.

\section{Discussion}

To characterize the ``chaotic probability'' observed in the nanoscale optical energy transfer between QDs~\cite{nanochaos}, we investigated one-dimensional random walks using fixed, periodic, quasiperiodic, and chaotic probabilities in this study.
The ETMSDs exhibited rich behaviors, although the overall time behaviors are classified as normal diffusion.
In particular, the ETMSD undergoes many large changes in weakly chaotic random walks, which we call ``accelerated diffusion,'' owing to the transient nonstationarity of its biases.

It is noted that the rapid increase in the ETMSD of No. 180 at the initial stage is only observed in the situation where we treat the time series as ``probabilities (thresholds).''
This rapid increase implies a short-time arrival to some points ($\pm$ $x$), as shown in Fig.~\ref{fig:3}.
In real physical situations, nanophotonic oscillations are the oscillations of the existence probabilities of the optical energy in QDs but not a probability of the left flight of a particle.
However, we believe that the properties of the nonsymmetricity and nonstationarity of the chaotic probabilities could generate the short-time arrival as well.
This ``nonstationary quick search'' property could be used for physically solving demanding computational problems implemented in QD arrays, such as a maze that consists of QDs~\cite{PRB,Lang,QDM}.

This characterization will also provide new applications using nanoscale chaotic probabilities, such as efficient optical energy transfer used for information propagation, efficient decision-making devices, and high-quality and high-speed physical nanoscale RNGs for information and communications technologies.

\section*{Acknowledgement}

This work was supported in part by the Core-to-Core Program, A. Advanced Research Networks
from the Japan Society for the Promotion of Science.
We thank Prof. Takahisa Harayama for valuable discussions.

\end{document}